%% file: 2019-wsdm-fp-explain-recs.tex
\documentclass[sigconf]{acmart}

\usepackage{latexsym}
\usepackage{graphicx}
\graphicspath{{./images/}}
\usepackage{booktabs} 
\usepackage{color}  
\usepackage{amsmath}  
\usepackage{subcaption}
\usepackage{caption}
\usepackage{balance}
\usepackage{verbatim}
\usepackage{cancel}
\usepackage{tikz}
\usepackage{colortbl} 
\usepackage{framed}
\usepackage{multirow}
\usepackage{multicol}
\usepackage{mathtools}

\theoremstyle{definition}
\newtheorem{definition}{Definition}[section]

\acmConference[WSDM '19]{The Twelfth ACM International Conference on Web Search and Data Mining}{February 11--15, 2019}{Melbourne, VIC, Australia}

\begin{document}

\title{FAIRY: A Framework for Understanding Relationships between Users' 
Actions
and their Social Feeds}

\author{Azin Ghazimatin}
\affiliation{%
  \institution{Max Planck institute for Informatics \\ Saarland Informatics Campus}
}
\email{aghazima@mpi-inf.mpg.de}

\author{Rishiraj Saha Roy}
\affiliation{%
	\institution{Max Planck institute for Informatics \\ Saarland Informatics Campus}
}
\email{rishiraj@mpi-inf.mpg.de}

\author{Gerhard Weikum}
\affiliation{%
	\institution{Max Planck institute for Informatics \\ Saarland Informatics Campus}
}
\email{weikum@mpi-inf.mpg.de}

\renewcommand{\shortauthors}{Ghazimatin et al.}

\newcommand\BibTeX{B{\sc ib}\TeX}

\newcommand{\struct}[1]{\texttt{\small #1}}
\newcommand{\utterance}[1]{\textit{``#1''}}
\newcommand{\phrase}[1]{\textit{`#1'}}
\newcommand{\old}[1]{}

\newcommand{\squishlist}{
 \begin{list}{$\bullet$}
  { \setlength{\itemsep}{0pt}
     \setlength{\parsep}{3pt}
     \setlength{\topsep}{3pt}
     \setlength{\partopsep}{0pt}
     \setlength{\leftmargin}{1.5em}
     \setlength{\labelwidth}{1em}
     \setlength{\labelsep}{0.5em} } }

\newcommand{\squishend}{
  \end{list}  }

\newcommand{\comm}[1]{}

\input{sections/abstract}

%
%
%
%

\maketitle

\input{sections/introduction}
\input{sections/model}
\input{sections/user-study}
\input{sections/experiments}
\input{sections/results}
\input{sections/related-work}
\input{sections/conclusion}


\bibliographystyle{ACM-Reference-Format}
\bibliography{explain-recs}

\end{document}

%% file: sections/abstract.tex
\begin{abstract}

Users increasingly rely on social media feeds for consuming daily information. The items in a feed, such as news, questions, songs, etc., usually result from the complex interplay of a user's social contacts, her interests and her actions on the platform. The relationship of the user's own behavior and the received feed is often puzzling, and many users would like to have a clear explanation on why certain items were shown to them. Transparency and explainability are key concerns in the modern world of cognitive overload, filter bubbles, user tracking, and privacy risks. This paper presents FAIRY, a framework that systematically discovers, ranks, and explains relationships between users' actions and items in their social media feeds. We model the user's local neighborhood on the platform as an interaction graph, a form of heterogeneous information network constructed solely from information that is easily accessible to the concerned user. We posit that paths in this interaction graph connecting the user and her feed items can act as pertinent explanations for the user. These paths are scored with a learning-to-rank model that captures relevance and surprisal. User studies on two social platforms demonstrate the practical viability and user benefits of the FAIRY method.
\end{abstract}

%% file: sections/introduction.tex
\section{Introduction}
\label{sec:intro}

\textbf{Motivation.} Web users interact with a huge volume of content every 
day, be it 
for news, entertainment, or inside social conversations.
To save time and effort, users are progressively depending on curated 
\textit{feeds} 
for such content.
A feed is a 
stream of \textit{individualized} 
content items
that a service 
provider tailors 
to a user.
Well-known 
kinds of
feeds 
include Facebook and Twitter for social networks, Quora and StackExchange 
for community question-answering, Spotify and Last.fm for music, Google News 
and Mashable for news, and so on. 
Since 
a
feed is 
a one-stop source
for information, it is 
important that users understand \textit{how items in
their feed relate to their profile and activity on the platform}. 

On some 
platforms like Twitter and Tumblr, the feed originates solely from updates in 
the user's social neighborhood or
from their explicitly stated interest categories, and the connection is almost 
always obvious 
to the user.
However, as service providers 
gather an increasing amount of 
user-specific information in an attempt to better cater to
personal preferences, more and more platforms (like Quora, LinkedIn, and 
Last.fm), are generating 
complex feeds.
Here, a feed results from
an intricate combination of one's interests, friendship network, her actions on 
the platform, and external trends.
Such platforms are the focus of this paper.
Over time, a user accumulates several thousands of 
actions that
together
constitute
her profile 
(posts, upvotes, likes, comments, etc.),
making it impossible for 
the user to remember all these details.
Further, the user may not even possess a complete 
record of 
her
actions on the platform,
a 
common situation
that has been referred to as the problem of 
\textit{inverse privacy}~\cite{gurevich2016inverse}.

In such situations, identifying
{\em explanatory relationships} between the users' online behavior
(social network, thematic interests, actions like clicks and votes) 
and the feed items they receive,
is useful
for at least three reasons:
(i) they can \textit{convince} the user of their relevance, whenever
a received item's connection to the user is non-obvious or surprising;
(ii) they can point the user towards future actionability (a course of action 
to 
avoid seeing more of certain kinds of items), and 
(iii) due to the sheer scale and complexity of data and models
that service providers deal with,
it is not always realistically possible to show end users why
some feed item was shown
to them; in such cases, these relationships are a proxy that the users could
find \textit{plausible}.

For example, if \textit{Alice} sees a post on making bombs in her feed when
she herself is unaware of any explicit connection to such, she might
be highly curious as to what she might have done to create such
an association. In this context, Alice would definitely find it useful
if she is now shown explanations like
the following, that could remind her of some relevant actions:
: (i) her good \textit{friend}
\textit{Bob}
is a close
\textit{friend} of \textit{Charlie}, who \textit{follows}
\textit{Chemistry}
and the \textit{bomb} post was \textit{tagged} as 
belonging to this category, or, (ii) she recently \textit{asked a question}
about
\textit{food}, that is recorded as a \textit{sub-category of}
\textit{Organics} in the platform's
taxonomy, and the author of the \textit{bomb} post has also
\textit{categorized} the post under \textit{Organics}.
In our study
involving $20$ users each on two platforms, participants reported seeing 
$2,410$ 
such non-obvious items in their feeds over a period of two months.


\textbf{Limitations of state-of-the-art.} In principle, service providers are
 in the best position to offer such 
explanations. But they rarely do so in practice. For example, Quora simply tags 
items not emanating directly from one's neighborhood or interest profile with 
\phrase{Topic you might like}, and Last.fm often has notes like
\phrase{Similar to Shayne Ward} for a track recommendation, neither elaborating 
the user's 
relationship to the artist Shayne Ward, nor how the similarity was determined
in the first place.
Facebook's explanations have similarly been brought into
question~\cite{andreou2018investigating}.
Except ~\cite{eslami2015always}, there is very little work investigating 
relationships between user interactions on the platform and 
items in their feeds (outside of Twitter~\cite{edwards2014bot},
where the feed is exclusively from the network).
%
Relationship discovery~\cite{liang2016links}, however, has been explored in 
other contexts and for 
different goals,
most notably for understanding 
entity relatedness
in knowledge graphs
~\cite{fang2011rex,bianchi2017actively,pirro2015explaining,seufert2016espresso,
lao2011random},
predicting links on social networks~\cite{zhang2014meta},
 and
generating personalized 
recommendations~\cite{yu2014personalized,lao2010relational}.

Understanding connections between user preferences and online advertisements
has been investigated in simulated environments to some extent
~\cite{lecuyer2014xray,lecuyer2015sunlight,parra2017myadchoices}.
We differentiate ourselves from these approaches in the following ways: 
\squishlist
\item Prior works have aimed
to
discover the \textit{true provenance} of an item using models of 
\textit{causality}, and are typically aimed at \textit{reverse-engineering}
the platform. 
This is 
very different
from our 
goal
where we try to unravel connections between
a \textit{user's own actions} and what she sees in her feed, to enable her
to better understand the interplay between her and the platform, and
for her to have a better handle on the cognitive overload resulting
from interactions with the platform;
\item A common approach in this setting is to use \textit{what-if} analysis
 methods~\cite{lecuyer2014xray,lecuyer2015sunlight} like
differential correlation, which are 
intractable in a setting
where 
a
user makes thousands of interactions with the platform over 
an extended period of time;
\item Mechanisms underlying advertisement targeting are guided by completely 
different (financial) incentives in comparison to regular feed items.
\squishend

\begin{figure} [t]
  \centering
   \includegraphics[width=1.0\columnwidth]{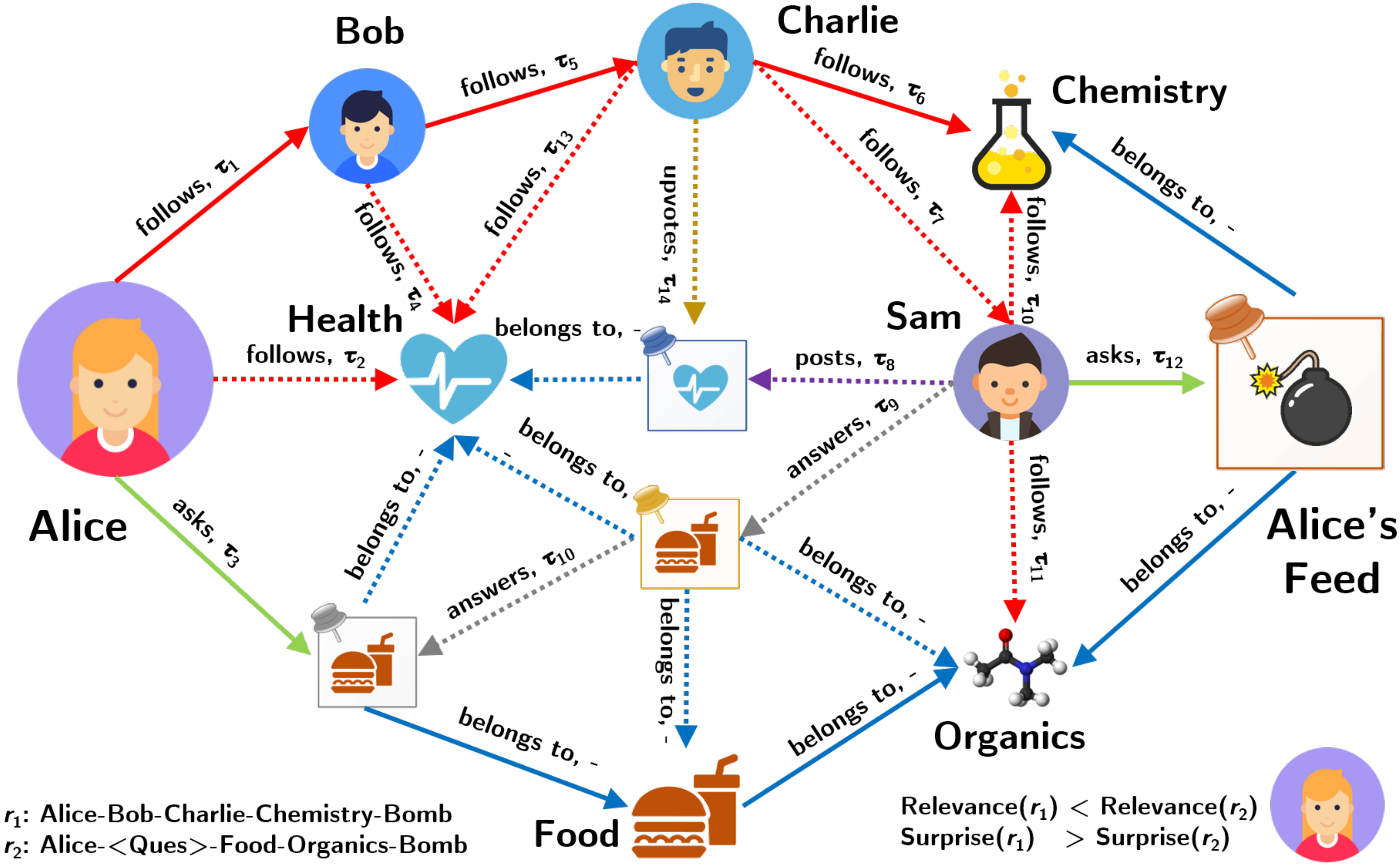}
      \caption{Toy interaction graph for Quora user Alice.}
      \label{fig:interaction-graph}
      \vspace*{-0.5cm}
\end{figure}
%

%


\textbf{Approach.} We propose FAIRY, 
a
\underline{\textit{F}}ramework for 
\underline{\textit{A}}ctivity-\underline{\textit{I}}tem 
\underline{\textit{R}}elationship 
discover\underline{\textit{Y}}, 
that
(i) 
addresses
the discussed challenges by building user-specific
\textit{interaction graphs} exclusively using information visible to the user
herself,
(ii) learns models 
for predicting relevance and surprisal,
trained on data from real user-studies on two 
platforms (Quora and Last.fm),
and
(iii) uses learning-to-rank techniques
to discover and rank relationships derived from the 
above interaction graphs.

Since our goal is to pinpoint a set of user actions and
their subsequent associations to the recommended feed item, 
FAIRY starts out by building an interaction graph connecting items in the
user's local neighborhood.
An example of such an interaction graph for a Quora user is shown in
Fig.~\ref{fig:interaction-graph}.
The interaction graph is modeled as a 
\textit{heterogeneous information network (HIN)}
~\cite{sun2013mining,sun2011pathsim,deng2011probabilistic,liang2016links}.
The HIN is a graph with
different types of nodes (users $\{u_l\}$, 
categories $\{c_m\}$, and items $\{i_n\}$) and edges (corresponding to 
different action 
types like \textit{follow, ask,} and \textit{upvote}, shown in different
colors). Nodes and edges in this HIN are weighted (not marked in 
Fig.~\ref{fig:interaction-graph} for simplicity).
Edges are directed and
\textit{timestamped}, 
corresponding to the time an action was performed, wherever applicable
($\tau_j$'s in Fig.~\ref{fig:interaction-graph}).
In the FAIRY framework, 
{each path in this interaction graph 
that 
connects the user $u$ (Alice) to a feed item $f$ (post on bombs)
corresponds to a 
potential
explanation 
for that
item, provided that each edge $e$ on the path has
timestamp $\tau(e) < \tau(f)$, where $\tau(f)$
is the time the feed item was seen by $u$}.
Two 
possible
explanation paths 
(out of many more)
are shown via solid edges in Fig.~\ref{fig:interaction-graph}.

The high 
number of
such explanation paths in the HIN demands
a subsequent ranking module.
We employ a learning-to-rank (LTR) model based on
ordinal regression~\cite{joachims2006training}, that 
models judgments of relevance and surprisal, collected from the \textit{same} 
users who 
received the feed item.
Features used in the LTR models
correspond to 
light-weight 
estimations of user influence, category specificity,
item engagement, path pattern frequency, and so on.
These are derived from node and edge weights in the HIN,
and are intentionally 
kept simple, to enable interpretability by the
end-user. 

We 
compare FAIRY to
three baselines: ESPRESSO~\cite{seufert2016espresso},
REX~\cite{fang2011rex}, and PRA~\cite{lao2010relational},
based on different underlying algorithms. These are 
state-of-the-art methods for
computing entity relatedness over knowledge graphs. 
FAIRY outperforms these
baselines on two representative platforms in modeling both relevance and
surprisal. Code for FAIRY is available at \url{
https://github.com/azinmatin/fairy}.

\textbf{Contributions.} 
This paper's key contributions are:
\squishlist
 \item the first 
user-centric
framework for discovering and ranking
 explanation paths between users' activities on a social network and items in 
 their feed;
 \item models for capturing the subtle 
aspects
 of user-relevance and surprisal for such explanations, 
with learning-to-rank techniques over light-weight features;
 \item extensive experiments including ablation studies, 
showing systematic 
 improvements over multiple
 baselines, and identifying vital factors in such models;
 \item a
user-study conducted over two months involving $20$ users each 
 on Quora and Last.fm, providing useful design guidelines for future research
 on feed analysis.
\squishend

%% file: sections/model.tex
\section{Discovering Explanations} 
\label{sec:model}

The FAIRY framework uses a heterogeneous information network
(HIN)~\cite{sun2013mining,liang2016links,yu2014personalized}
to represent
a user's presence and activities on a social network $S$ as a graph, and
relies
on a learning-to-rank (LTR) model to order explanation paths mined from
this
user-specific interaction
graph.


We are given a user $u \in U$, where $U$ is the set of members of $S$, 
and her feed on $S$, $F_u^S$. Our goal here is to find and rank the set of
explanation relations $R_{uf} = \{r_{1}, ...., r_{k}\}$
between $u$ and $f \in F_u^S$, where $k = |R_{uf}|$.
$R_{uf}$ is the set of connections
between $u$ and $f$ via $u$'s local neighborhood on $S$.
This local neighborhood is initialized using the set of actions that
$u$ performs: $A_{u} = \{a_{1}, ..., a_{m}\}$, where $m = |A_u|$.
Examples of such activities are \phrase{asking a question} on Quora
or \phrase{loving an album} on Last.fm. Connections to $f$ via $A_u$ are
identified by traversing the vicinity of $u$ in $S$, and recorded
in $u$'s interaction graph, $G_u$. Strictly speaking, $G_u$ may not be
fully connected. Typically, however,
$S$ has an underlying taxonomy $\mathbb{T}$ of topics that all entities
in the network must belong to, and overlaying $\mathbb{T}$ on
the vicinities of $u$ and $f$ ensures connectivity in $G_u$.
Specifically, a path between any pair of entities may be found
by first associating them via the categories they directly belong to,
followed by subsequent generalization by traversing higher up in $\mathbb{T}$,
till a connection is established (see~\cite{kasneci2009star} for an
application of this strategy in relationship mining).
More formally, each $G_u$ is 
an instance of a HIN, defined as:

\begin{definition}
	\label{def:ig}
	The \textit{\textbf{Interaction Graph}} of a user $u$ is a directed 
	and weighted multi-graph $G_u = (N, E, M^T_N, M^T_E, M^W_N, M^W_E,
	M^\mathcal{T})$,
	where: $N$ is the set of nodes,
	$E$ is the set of edges,
	$M^T_N: N \mapsto T_N$ and $M^T_E: E \mapsto T_E$
	are functions mapping nodes and edges to their corresponding types
	(sets $T_N$ and $T_E$),
	$M^W_N: N \mapsto \mathbb{R}_{\geq 0}$ and 
	$M^W_E: E \mapsto \mathbb{R}_{\geq 0}$
	are node and edge weight mapping functions respectively, and
	$M^\mathcal{T}: E \mapsto \mathcal{T}$ maps each edge $e \in E$ to
	a timestamp $\tau(e)$.
\end{definition}

The user $u$, her actions $A_u$, and the feed item $f$ are naturally part
of $G_u$ ($u, f \in N, A_{u} \subseteq E$). Fig.~\ref{fig:interaction-graph}
shows a representative interaction graph where $u$ is the leftmost user Alice,
and $f$ is the rightmost item marked with a bomb. We now explain each property
of this HIN, instantiating them with corresponding features of social networks.

\textbf{Nodes} $\{n \in N\}$ in $G_u$ correspond to entities in
the social network.
\textbf{Node types} are either users or various classes of content
(categories, tags, posts, songs,
etc.). Via mapping $M^T_N$, we have $M^T_N(n) = t_N^n$, where
$t_N^n \in T^N$ is one of the types above for node $n$.
In Fig.~\ref{fig:interaction-graph},
$N = \{\textrm{Alice, } \ldots, \textrm{Health, }, \ldots,
\textrm{bomb-post} \}$, such that $M^T_N$(Alice) = user,
and $M^T_N$(Health) = category.

An \textbf{edge} $e \in E$ represents a relationship (interchangeably
referred to
as \textit{actions} henceforth) between two nodes $n_1, n_2 \in N$.
Edges represent the following connections:
(i) \textit{user-content}: these capture engagement or actions by the users on
the content in $S$ (in Fig.~\ref{fig:interaction-graph},
user Alice $\xrightarrow{follows}$ category Health),
(ii) \textit{user-user}: they capture social relationships between users
(Charlie $\xrightarrow{follows}$ Sam),
and (iii) \textit{content-content}:
these edges capture relationships between content items
(Food $\xrightarrow{belongs\;to}$ Organics).
Each $e$ is mapped to an \textbf{edge type}
$M_E^T(e) = t_E^e$,
where $t_E^e \in T_E$, instantiated depending on
the platform (e.g., \textit{asks, answers, follows,} $\ldots$ for Quora);
$t$(Charlie $\xrightarrow{follows}$ Sam) = follows).
For each edge in the HIN, we add
an opposite edge \textit{typed} with the \textit{inverse relation} between
the same pair
of nodes (e.g., we add Health $\xrightarrow{follows^{-1}}$ Alice).
This enables \textit{bi-directional traversal} of the HIN.

Nodes $n \in N$ and edges $e \in E$ in $G_u$ are associated with
non-negative \textit{weights}.
The \textbf{node weight} $M_N^W(n)$ may reflect
node \textit{influence, specificity, or engagement} depending upon
the entity type. The value of $M_N^W(n)$
itself may be derived from measurable features of the platform that
are visible to $u$. For example, if the post on food was upvoted by $30$ 
users,
$M_N^W(\textrm{food-post}) = 30$. An \textbf{edge weight} $M_E^W(e)$ counts
the number of times the action
was performed, e.g., a Last.fm user can \textit{scrobble} (listen to)
a specific song five times ($M_E^W(e) = 5$).

$G_u$ is a \textbf{multi-graph}, implying that there may exist more than
one edge
between any two nodes, corresponding to multiple actions. For example, Twitter
users can both \textit{like} and
\textit{re-tweet} a Tweet, and users on Last.fm can both \textit{scrobble}
a song, and \textit{love} it.

There exists a unique \textbf{edge timestamp} denoted by $\tau(e)$,
which is the time when the corresponding action was performed
(e.g., $\tau(\textrm{Bob} \xrightarrow{follows} \textrm{Charlie}) = \tau_5$).
This is possibly null, if $e$ has existed since the epoch
(category memberships are assumed to be such edges in this work).
For edges with
$w(e) > 1$ (action performed multiple times), we define
$\tau(e)$
as the timestamp of the first instance of this action.
 

The size of the interaction graph $G_u$ is characterized by
the \textit{eccentricity} of $u$, $\epsilon(u)$. This is the greatest
\textit{graph (geodesic) distance}
(length of shortest path) $d$ between $u$ and any other node $n'$
in $G_u$, i.e.,
$\epsilon(u) = \max_{n' \in G, n' \neq u} d(u, n')$.
In other words, $\epsilon(u) = 3$ yields the $3$-hop neighborhood of $u$
in $G_u$. In Fig.~\ref{fig:interaction-graph},
$\epsilon(\textrm{Alice}) = 4$.


We are now in a position to discover
the set of explanation relations $R_{uf}$, formally defined below.

\begin{definition}
	\label{def:valid-path}
	An \textbf{\textit{Explanation Path}}
	is a path $r$ connecting $u$ and $f$ in $G_u$
	such that the timestamp of every edge $\tau(e)$ in 
	$r$ is strictly less than the time when $f$ was seen by $u$,
	i.e. $\tau(f) > \max_{e \in r} \tau(e)$.
\end{definition}

 

In Fig.~\ref{fig:interaction-graph}, 
for
(Alice, bomb-post $f$), if we have $\tau(\textrm{bomb-post}) = 13$,
then $\textrm{Alice}
\xrightarrow{follows} \textrm{Health}
\xrightarrow{belongs\;to^{-1}} \textrm{health-post} \xrightarrow{posts^{-1}} 
\textrm{Sam} 
\xrightarrow{asks}\textrm{bomb-post}$, is a valid explanation path.
However, the following path:
$\textrm{Alice} \xrightarrow{follows} \textrm{Bob}
\xrightarrow{follows} \textrm{Charlie} \xrightarrow{upvotes} 
\textrm{health-post} \xrightarrow{posts^{-1}}
\textrm{Sam} \xrightarrow{asks}
\textrm{bomb-post}$, is \textit{invalid}
as the timestamp $\tau(\textrm{Charlie} \xrightarrow{upvotes} \textrm{health-post})
 = 14 > 13$.
Henceforth, \textit{explanations} and \textit{relations}
(with or without \textit{paths}) are used interchangeably, and should be
understood as equivalent.
An explanation path can thus be a combination of user-content,
user-user, and content-content edges. For example, the path outlined
earlier in this paragraph is a mixture of user-content and content-content
edges, but lacks any user-user connection. Thus, given, 
$u$, $f$, and $G_u$,
we extract all explanation paths between $u$ and $f$ from $G_u$
as candidates for further processing.

It can be argued that since the interaction graph could often be dense, 
explanations could be better
presented as \textit{sub-graphs}~\cite{seufert2016espresso} instead.
But we prefer paths in this work due to
the following reasons: (i) sub-graphs are difficult to isolate by influence,
especially for dense neighborhoods; (ii) paths (simple, without loops),
are atomic units of relationship, and subgraphs can, in fact, be reduced
to a number of constituent paths; and (iii) subgraphs are harder to
interpret for the average user, and may be more difficult to make
comparative assessments. 

\section{Ranking Explanations}
\label{sec:ranking}

Due to the high activity count aggregated by a user over her time as a member
of the platform, and due to the richness of the platform itself (allowing
a post to have multiple categories, having a detailed directed acyclic
graph (DAG) taxonomy, etc.), the usual number of candidate explanation paths
is too high to be processed by a user, if presented all at once. Measurements
from our user study show that the number of such paths can vary from a few
thousand to even millions (depending on the graph size
determined by $\epsilon(u)$, and length of relation path $r$). Thus, it is
imperative that we rank these paths and present only a top few, to prevent
a cognitive overload.

Over the last decade or so,
learning-to-rank (LTR)~\cite{cao2007learning,joachims2006training}
as emerged as
the \textit{de facto}
framework for supervised ranking in information retrieval and data mining,
which motivates its application to 
FAIRY.
LTR has three basic
variants: pointwise, pairwise,
and listwise, with
each type proving beneficial in specific contexts
~\cite{bast2015more,radlinski2005query}.

The common guiding criterion, though, is the nature of gold label judgments
that can be collected (and/or inferred). In our case,
we want to model 
relevance and surprisal of the explanation paths. Generally, it is difficult for
a user to score a standalone explanation path on scale of $0-10$, say.
At the other
extreme, it might be even harder to score a complete list of say,
a heuristically chosen sample of ten paths. Collecting
\textit{preference judgments} is the most natural thing to do in our setting:
it is
a conceivable task for an end-user (an average social media user, here)
to rate a path as being \textit{more relevant} (generally useful as
a satisfactory explanation to her),
or it being \textit{more surprising} (such as discovering
a forgotten/unknown connection
in her vicinity),
\textit{than another path} connecting her to the same feed item. Collecting
such explicit
pairwise annotations has been suggested as being cognitively preferable to
the user in other contexts like
document relevance assessments~\cite{carterette2008here}.
In a similar vein, we use a pairwise learning-to-rank model based on
ordinal regression, that directly makes use of the users' preference judgments
on pairs of explanations~\cite{joachims2006training}. We used $SVM^{Rank}$
with a linear kernel, as it is very fast to train,
and has been shown
to be highly effective in ranking result pages for
search queries~\cite{schuhmacher2015ranking}.


\subsection{LTR features}
\label{subsec:feat}

The general principles guiding this work
are explainability and transparency. This influences the choice of our features
in two ways: (i) while the provided explanation should already be insightful
to the user, it is not unreasonable to assume that the user could, in turn,
want to know what was responsible for a few chosen paths to be shown as
more surprising or relevant than others. This points towards using
simple and interpretable features that can give a na\"{i}ve user a handle on
what was found to
be ``important'' in this context; and, (ii) the features should
be \textit{visible}
to the user in question: either public, or easily accessible in a few clicks,
or by visiting a user-specific URL. Data that only the service provider
has access to (derived measures of user-user similarity), or requires 
excessive crawling (total number of authors of all posts in a category) are
clearly unsuitable in the task at hand. With these guidelines, we define
the following sets of features for the FAIRY framework. These are grouped
into five sets: (i) user, (ii) category, (iii) item, (iv) path instance,
and (v) path pattern. For the first three sets, if there are multiple instances
of the same type on a path (two users or three categories), the feature
value is averaged over these instances.

\subsubsection{User features}
\label{subsubsec:user}

We consider two factors for users on
explanation paths: (i) \textit{user influence}, and (ii) \textit{user activity}.
Influence is typically measured using number of followers or the link ratio
(ratio of followers to followees, wherever \textit{friendship}
is not mutual)
on social networks~\cite{srijith2017longitudinal}. Higher the link ratio,
higher is the perceived influence. For activity, we
measure the individual types of activity that the user is allowed on 
the platform (\textit{scrobbling} tracks, \textit{loving} tracks,
and \textit{following} other users on Last.fm). More influential and active
users may have a discriminative effect on the user's judgments for relevance.
	
\subsubsection{Category features}
\label{subsubsec:cat}

(i) \textit{Popularity} or influence can be
estimated
for categories too, for example, by counting the number of posts in them,
or looking at their total numbers of followers or subscribers. Such aggregates
are often made visible by providers, and it is not necessary to actually
visit and count the items concerned. (ii) We also consider
\textit{category specificity}~\cite{ramakrishnan2005discovering},
which is reflected by its depth in the category hierarchy
(the higher the level, the more specific it is, root assumed to be level $0$)
and the number of children (sub-categories) it has in the taxonomy
(more children implying less specificity).
While popularity may influence user's inputs similarly as for users,
high category specificity may directly affect the surprisal factor.

\subsubsection{Item features}
\label{subsubsec:item}

(i) \textit{Specificity} for items (songs, posts, etc.)
may be analogously computed 
by counting the number of different categories that the item belongs to.
(ii) \textit{Engagement} is an important measure for items, which represents
the different
actions that have been performed on it (typically the same as the number
of users
who have interacted with the item). Examples include the number of different
listeners of a song on Last.fm, or the number of different answers that a 
question has on Quora, etc. Engagement may be perceived as analogous to 
influence or popularity for users and categories in its role in FAIRY.
	
\subsubsection{Path instance features}
\label{subsubsec:path-inst}

There are some properties of the explanation
path as a whole: these can be understood better by further separating them
into path instance and path pattern features. Instance features measure
aspects of the specific path in question. These include:
(i) Aggregate
\textit{similarity of feed item $f$ to relation path $r$},
i.e.,
$sim(f, r) = \frac{\sum_{n \in r, n \neq u, f} sim(n, f)} {len(r) - 1}$,
where $n$ is any \textit{internal node} on the path, and $len(r)$ is
the path length ($-1$ gives the number of such internal nodes).
This normalized similarity function $sim(\cdot, \cdot) \in [0, 1]$
is treated as a plug-in and can be instantiated
via embedding-based similarities, or computed locally from $G_u$ itself.
The item-path similarity function, in a sense, measures
the \textit{coherence}
of the extracted relation path to the recommended item. If it is very low,
the path could be quite surprising to the user.
(ii) Analogous aggregate \textit{similarity of user $u$ with $r$}:
$sim(u, r) = \frac{\sum_{n \in r, n \neq u, f} sim(n, u)} {len(r) - 1}$.
This feature models the familiarity of the user with the path as a whole.
(iii) \textit{Path length} $len(r)$: The length of an explanation is easily
one of the most tangible factors for a user to make decisions on:
while shorter paths may imply obvious connections, longer paths may be
more surprising.
(iv) \textit{Path recency}: This is a temporal feature, and is defined as the
most recent edge on the path with respect to the feed:
$\min_{e \in r} \tau(f) - \tau(e)$. Highly recent paths will have a low
value of this feature; the idea is that relatively newer paths may be more
relevant to the user due to freshness, while older paths may have
an associated surprise factor.
(v) \textit{Edge weights}: This feature averages the number of times 
each action on the path has been repeated (e.g. number of times a user 
has listened to a specific song). Note that this feature cannot be used 
for Quora as none of the actions can be repeated, i.e, the user cannot 
upvote/follow/ask/answer the same item more than once. 

\subsubsection{Path pattern features}
\label{subsubsec:path-patt}

Pattern features drop concrete instantiations $\{n\}$ and $\{e\}$
in $r$ and deal with
the underlying sequence of node and edge types $M^T_N(n)$ and $M^T_E(e)$
instead
($r_1$ in
Fig.~\ref{fig:interaction-graph} has the path pattern:
$user \xrightarrow{follows} user \xrightarrow{follows} user
\xrightarrow{follows} category \xrightarrow{belongs\;to^{-1}} post$).
We consider the following features:
(i) \textit{Pattern frequency}: This is the average 
\textit{support} count
of a pattern between any $u$ and $f$ 
(frequent patterns may be less surprising).
(ii) \textit{Pattern confidence}: The percentage of $(u, f)$ pairs 
with at least one observed instance of the pattern.
(iii) \textit{Edge type counts}: 
Users in our study explicitly mentioned the effect of some edge types on 
their choice of relevance and suprisal.
Thus, to zoom into the effect produced by
the aggregate measure of pattern frequency, 
we also considered the counts of each 
edge (action) type
present on the path as features 
(\#likes, \#\textit{follows}, \#\textit{belongs to}, etc).

The general intuition here is that the user has clear mental models
of relevance and surprisal; the above features are what she sees in her daily
interactions with the platform. These are the tangible perceptions
that influence her models, and the aim here is to \textit{learn}
how these factors combine to mimic her assessments.


%% file: sections/user-study.tex
\section{User studies}
\label{sec:user-study}

\begin{figure} [t]
    \centering
    \begin{subfigure}[b]{0.5\columnwidth}
        \includegraphics[width=\columnwidth]{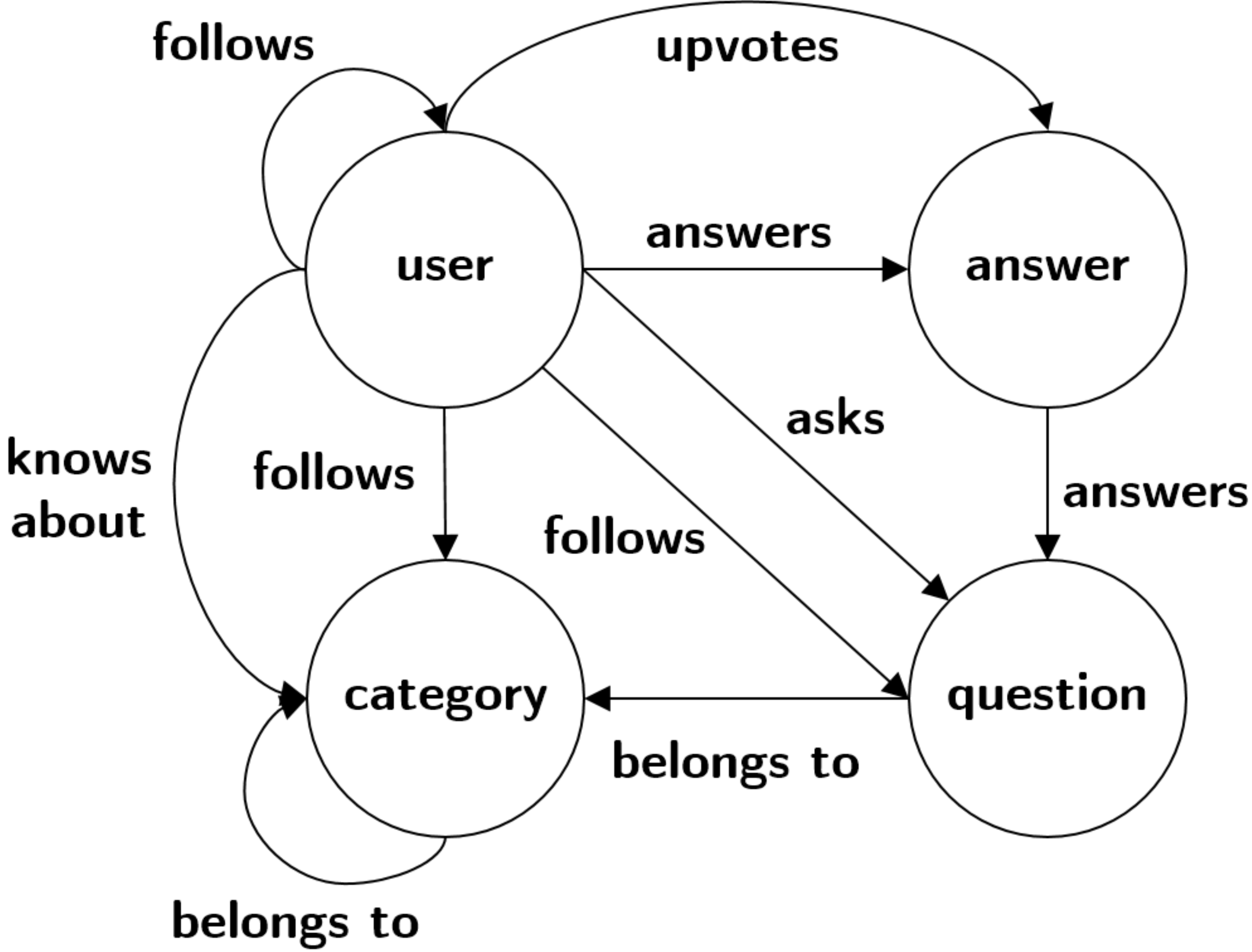}
        \caption{Quora}
        \label{fig:quora-schema}
    \end{subfigure}
    ~ 
    \begin{subfigure}[b]{0.5\columnwidth}
        \includegraphics[width=\columnwidth]{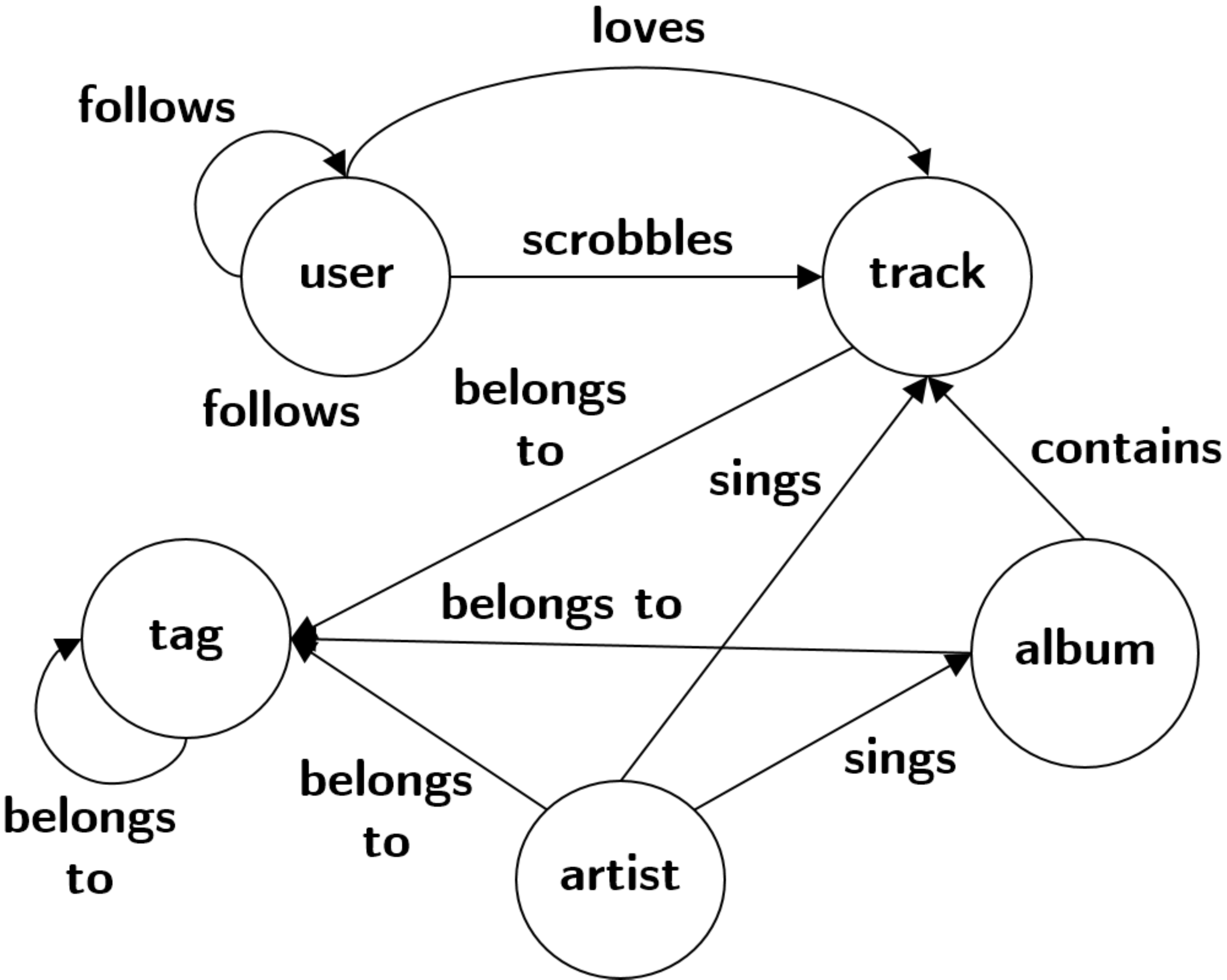}
        \caption{Last.fm}
        \label{fig:lastfm-schema}
    \end{subfigure} 
    \caption{Logical schemata for the Quora and Last.fm platforms,
    showing permissible relationships between node (entity) and
    edge (action) types.}
    \label{fig:schemas}
\end{figure}

\textbf{Platforms.} Since user feeds are never public, we needed to design 
user studies from where
we could collect gold judgments on explanation paths extracted and ranked
by FAIRY.
While there are a plethora of social network platforms providing
personalized feeds, we chose Quora
and Last.fm as they possess the richness
that can truly test the full power of our HIN model. To be specific,
these platforms
have node types, edge types, non-obvious feed items, properties for estimating 
node and edge weights, have millions of members, and have been subject
of previous research (e.g., Quora in ~\cite{wang2013wisdom},
Last.fm in ~\cite{jaschke2007tag}). Also, one being
a community-question answering site, and the other an online
music recommender, they represent two completely different application
platforms, and hence ideal to evaluate FAIRY with. We use a slightly
simplified version of these platforms' schemas, as shown in
Fig.~\ref{fig:schemas}.

\textbf{Users.} We hired $20$ users each for Quora and Last.fm, for interacting
with the
platforms and assessing explanations, in May - July 2018. Users had to
set up fresh accounts
(using real credentials for accountability) so that all their activity
can be recorded. Each user
had to spend $20$ hours on one platform in total. This total time was divided
into one hour sessions per day, with a gap of one day between consecutive
sessions. Interactions were planned in this staged manner so as to allow
the service provider enough time to build up user profiles, and generate
personalized feeds. The hired users were graduate students of mixed background
who were familiar with Quora and Last.fm. They were paid \$10 per
$1$-hour session, for three tasks: (i) interacting with the platform
(a minimum of $12$ activities per session from permissible actions
in Fig.~\ref{fig:schemas}, no upper bound, ``natural'' behavior
recommended),
(ii) identifying non-obvious feed items after going through their 
complete feed (both platforms have a countable feed at a given point of time), 
 and (iii) providing assessments of
relevance and surprisal. Their complete activity and feed on the platforms
were recorded.
Users were given a random set of ten initial topics to follow on Quora,
due to such a requirement by the platform. They were assigned five
initial followers from within the study group on Quora and Last.fm.

\textbf{Judgments.} After each session, we updated the interaction graphs of users,
selected three non-obvious recommendations per user, and mined explanation
paths for these feed items. Path length was restricted to four for Quora
and five in Last.fm, which resulted in about $2k-30k$ paths on average
over $(u, f)$ pairs. Increasing path length led to
an \textit{exponential increase}
in the number of explanations. About $25$ pairs of paths were then
randomly sampled
per feed item, to be assessed by the user
on two questions:
(i) Which explanation path is more relevant (useful) to you for this feed item?
(ii) Which explanation path is more surprising to you for this feed item? 
The users were allowed to give free-text comments on reasons motivating
their choices. Several of these intuitions and allusions were encoded into
our design considerations.

\textbf{Ethics.} To comply with standard ethical guidelines, all participants
were informed about the purpose of the study, and that their data was being
collected for research purposes. They signed documents to confirm their
awareness of the same. Users signed up with their real credentials; no terms
of service of the providers were violated over the course of the research.
All users deleted their accounts at the conclusion of the study.

%% file: sections/experiments.tex
\section{Evaluation Setup}
\label{sec:experiments}


\begin{table}[t] 
	\resizebox{\columnwidth}{!}{
	\begin{tabular}{l c c c c}
		\toprule
		\textbf{Platform} 	& \textbf{\#Activity} 	& \textbf{|\textit{N}|} & \textbf{|\textit{E}|} & \textbf{\#(\textit{u}, \textit{f}) paths} \\ \toprule
		Quora 				& $217$					& $31,769$ 				& $532,569$				& $28,527$									\\
		Last.fm 			& $132$ 				& $22,815$ 				& $79,252$ 				& $1,897$									\\ \bottomrule
	\end{tabular}}
	\caption{User study statistics (nos. averaged over users).}
	\label{tab:user-study-measurements}
	\vspace*{-1cm}
\end{table}



\noindent \textbf{Datasets}. For Quora, we used $11,677$ pairs of
explanation paths evaluated
by $20$ users as the 
gold standard. These explanation paths covered $459$ distinct $(u,f)$ pairs.
Each 
explanation path appeared in $1.9$ pairs on average.
For Last.fm, we collected $4,791$ evaluated pairs
from $20$ users. These paths 
were extracted as potential explanations for $235$ distinct $(u, f)$ pairs.
Each path occurred in about $1.7$ pairs.
Details on
the interaction graphs 
are in Table~\ref{tab:user-study-measurements}.

\noindent \textbf{LTR.} To run LTR, we divided each dataset into
$80\%$ training, $10\%$ development,
and $10\%$ test sets. We used $SVM^{rank}$~\cite{soh2013recommendation}
with a linear kernel in all our experiments.

\noindent \textbf{Baselines}. FAIRY was compared with three baselines for
relationship discovery: 
ESPRESSO~\cite{seufert2016espresso}, REX~\cite{fang2011rex} and
PRA~\cite{lao2010relational}. 
In ESPRESSO~\cite{seufert2016espresso}, the goal is to find
relatedness cores (dense subgraphs)
between two sets of 
query nodes. For this, they first identify a center, i.e, a node with
highest similarity to both 
the input query sets. Then, they expand the subgraph by adding other 
key entities, their context entities, and query context entities. 
In each step, the entities
are selected based 
on random-walk based scores. To apply this algorithm, we consider
$\{u\}$ and $\{f\}$ as the input query sets. For each path, we compute 
its score as if it were the output of ESPRESSO. 
For this, we first find the the most similar node on the path to both $u$ and $f$ as 
the center. We then expand the set of selected nodes by adding their 
adjacent nodes on the path. At the time of adding each node, we compute 
their random-walk based similarity to the adjacent (already selected) node on 
the path. At the end, we compute the score of each path by averaging 
the scores of its comprising nodes.  

REX~\cite{fang2011rex} takes a pair of entities and returns a ranked list of
its relationship explanations. 
Like ESPRESSO, the relationships are in the form of subgraphs.
The extracted relationships are then 
ranked based on different classes of measures. These measures can 
be used for paths as well. We used all aggregate and
distributional measures from the original paper.
However, for brevity, we only describe the global distributional measure as
it performed the best. This measure
captures rarity of relations as a signal of interestingness. For this,
we compared the value of $1-pattern\_confidence(r)$ for each
explanation path $r$.

For PRA~\cite{lao2010relational}, we
computed scores of paths via pattern-constrained random walks.
For instance, a pattern like ``user $\xrightarrow{follows}$ post''
only 
allows the random walker to leave the source node with type ``user''
to nodes with 
type ``post'' through edges of type ``\textit{follows}''.

\noindent \textbf{Metric.} We measured the percentage accuracy of each method
(or configuration, as applicable) as the ratio of correct predictions to
all predictions over pairs of relationship paths.

%% file: sections/results.tex
\vspace*{-0.1cm}
\section{Results and Insights}
\label{sec:results}

\subsection{Key findings}

\begin{table}
	\resizebox{\columnwidth}{!}{
	\begin{tabular}{c l c c c c c}
		\toprule
		\textbf{Platform} & \textbf{Method} & \textbf{FAIRY} & \textbf{ESPRESSO~\cite{seufert2016espresso}} & \textbf{REX~\cite{fang2011rex}} & \textbf{PRA~\cite{lao2010relational}} \\ \toprule
		\multirow{2}{*}{Quora} 		& Relevance & $\boldsymbol{60.33}$* & $49.93$ & $23.47$ & $30.84$	\\
									& Surprisal & $\boldsymbol{60.38}$* & $49.93$ & $21.58$ & $51.93$ 	\\ \midrule
		\multirow{2}{*}{Last.fm}	& Relevance & $\boldsymbol{56.24}$* & $48.85$ & $37.66$ & $49.06$ 	\\
									& Surprisal & $\boldsymbol{54.21}$* & $51.39$ & $36.03$ & $43.29$ 	\\ \bottomrule
	\end{tabular}}	
	\caption{Accuracy of FAIRY compared with baselines. The maximum value in a row is marked in bold. An asterisk (*) denotes statistical significance
	of FAIRY over the strongest baseline,
	with $p$-value $\leq 0.05$ for a two-tailed paired $t$-test.}
	\label{tab:baselines}
	\vspace*{-1cm}
\end{table}
 
 \begin{figure} [t]
 	\includegraphics[width=0.8\columnwidth]
 	{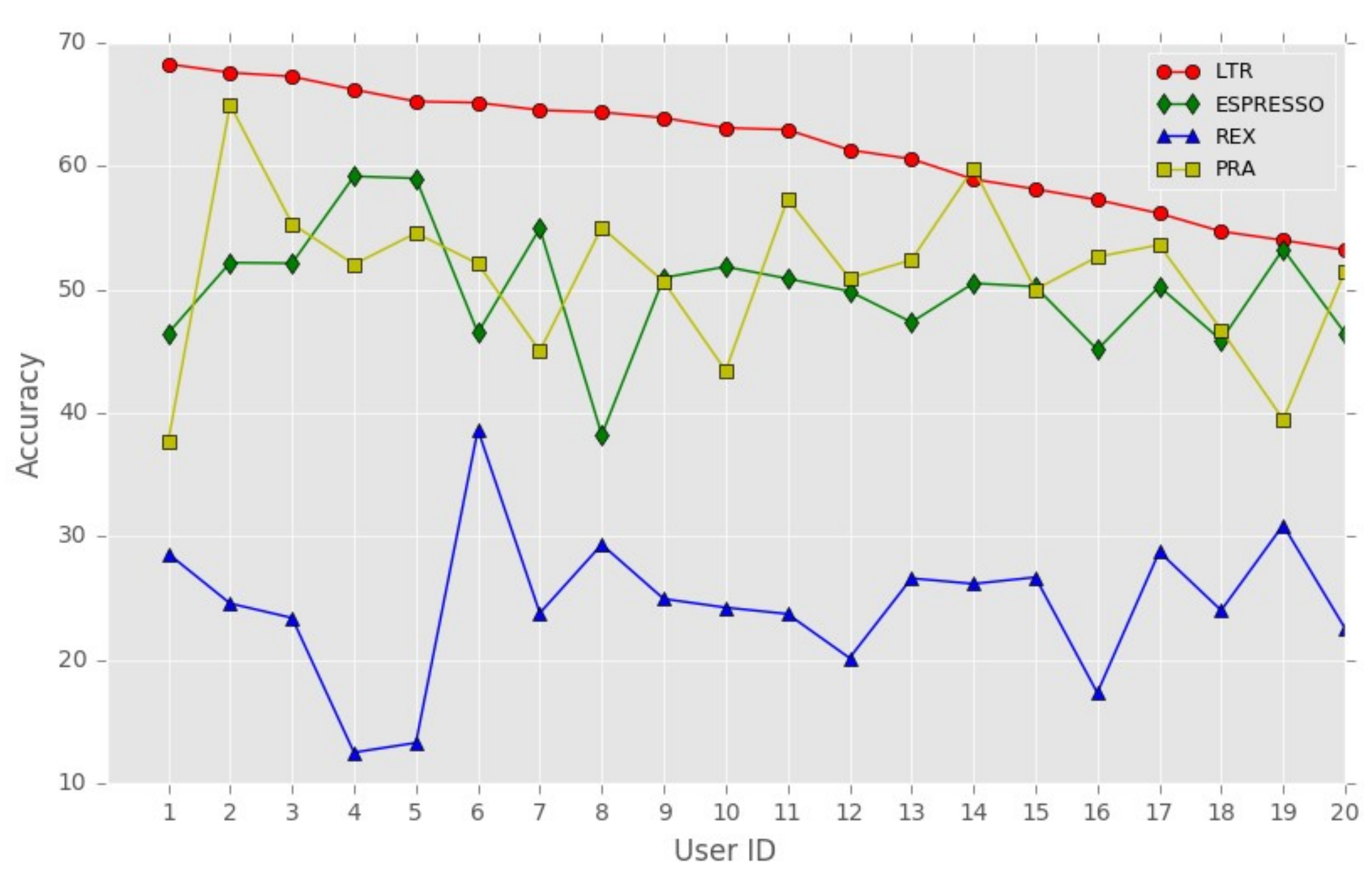}
 	\caption{User-specific models for relevance (Quora).}
 	\label{fig:user-quora-rel}
 	\vspace*{-0.5cm}
 \end{figure} 
 
\noindent \textbf{Comparison of FAIRY with baselines.} Table~\ref{tab:baselines} shows 
the comparison of accuracy for the relevance and surprisal models of FAIRY with 
baselines on both 
platforms. In all cases, FAIRY significantly outperforms all the baselines 
(paired $t$-test with $p$-value $<$ 0.05).
Note that all
baselines have the same model for both relevance and surprisal as they try to 
find either \phrase{relevant} or \phrase{interesting} relationships.
All baselines solely rely on structural properties of the underlying graphs.
In ESPRESSO, scores of cores are affected by degrees
of intermediate nodes. More precisely, on Quora, category 
nodes have large degrees as they are connected to many other nodes. 
This affects scores of paths with many intermediate category nodes.
A similar 
problem happens in PRA as it is also based on random walk. Besides,
as path 
scores are computed by multiplying reciprocals of node degrees, PRA is 
biased toward shorter paths.
In REX, we are only able to compare explanation paths with different 
path patterns. This has substantial effect on accuracy, as many
explanation paths share the same pattern. 

\begin{figure}[t]
 	\includegraphics[width=0.8\columnwidth]
 	{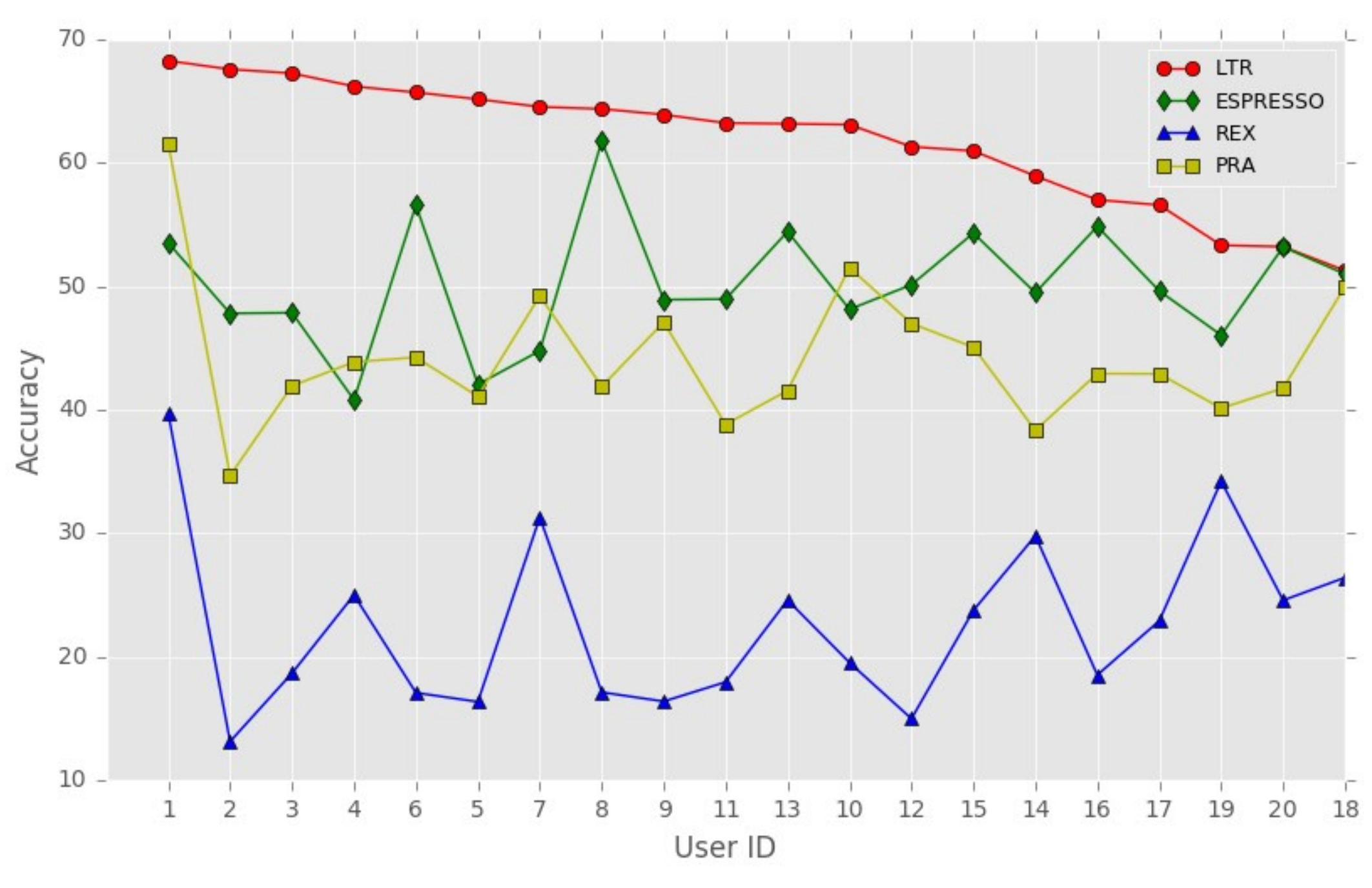}
 	\caption{User-specific models for surprisal (Quora).}
 	\label{fig:user-quora-sur}
 	\vspace*{-0.5cm}
\end{figure}

\begin{figure} [t]
 	\includegraphics[width=0.8\columnwidth]
 	{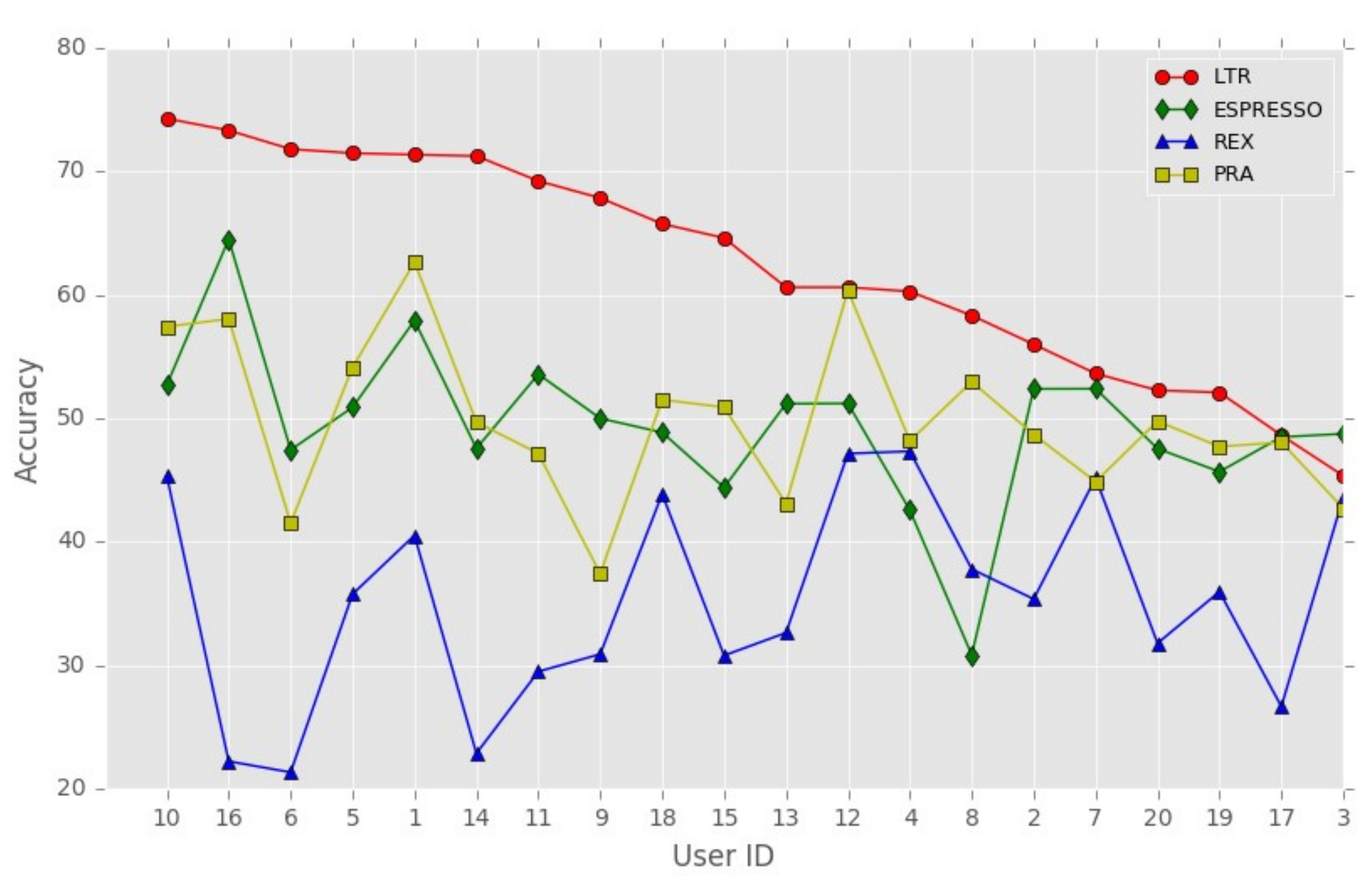}
 	\caption{User-specific models for relevance (Last.fm).}
 	\label{fig:user-lastfm-rel}
 \end{figure}

\begin{figure} [t]
 	\includegraphics[width=0.8\columnwidth]
 	{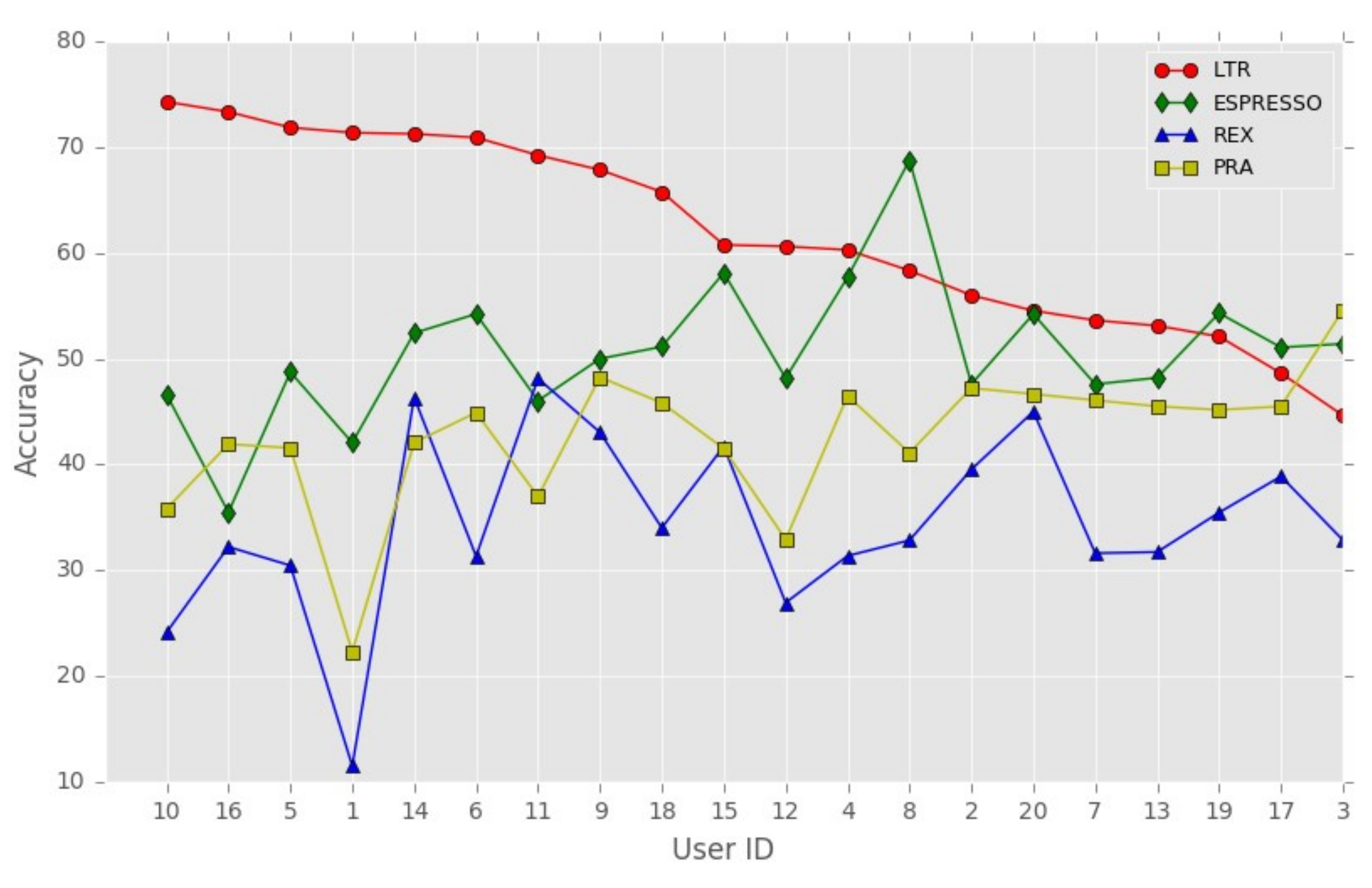}
 	\caption{User-specific models for surprisal (Last.fm).}
 	\label{fig:user-lastfm-sur}
 \end{figure}

\noindent \textbf{User-specific models.} To test subjective preferences 
for relevance and surprisal, we built and evaluated analogous
user-specific LTR models.
FAIRY accuracies of user-specific models were observed to be higher than
the aggregate global model for most users
(Fig.~\ref{fig:user-quora-rel}-\ref{fig:user-lastfm-sur}). There are, however,
a few users whose 
judgments we could not easily predict.
User id's are assigned by descending order of FAIRY accuracy
in Fig.~\ref{fig:user-quora-rel}, and these id's are used as references
for comparison across all subsequent figures.
We found that
FAIRY  
outperforms baselines in user models as well. Note,
again, that baselines 
do not have separate models for relevance and surprisal.

\subsection{Analysis and Discussion}

\begin{table} 
	\resizebox{\columnwidth}{!}{
		\begin{tabular}{l c c c c c}
			\toprule
			\multirow{2}{*}{\textbf{Features}} & \multicolumn{2}{c}{\textbf{Quora}} & & \multicolumn{2}{c}{\textbf{Last.fm}}	\\
			\cmidrule{2-3}\cmidrule{5-6} 
							& \textbf{Relevance} 	& \textbf{Surprisal} 	& & \textbf{Relevance} 		& \textbf{Surprisal} 	\\ \toprule
			All 			& $\boldsymbol{60.33}$ 	& $60.38$ 				& & $56.24$ 				& $54.21$ 				\\ \midrule
			No user 		& $60.21$ 				& $60.19$ 				& & $56.03$ 				& $54.15$ 				\\
			No category 	& $\boldsymbol{60.33}$ 	& $60.23$ 				& & $\boldsymbol{56.65}$ 	& $\boldsymbol{54.80}$ 	\\
			No item 		& $60.31$ 				& $60.38$ 				& & $56.24$ 				& $54.42$ 				\\
			No path patt. 	& $51.69$ 				& $51.73$ 				& & $54.32$ 				& $53.65$ 				\\
			No path inst. 	& $60.21$ 				& $\boldsymbol{60.71}$	& & $55.78$ 				& $54.21$ 				\\ \bottomrule
		\end{tabular}}
	\caption{Ablation study results. The highest value in each column is marked in bold.}
	\label{tab:ablation}
	\vspace*{-0.5cm}
\end{table}

\noindent \textbf{Ablation study}. To analyze effects of different
feature groups on LTR 
accuracy, we removed one group at a time and retrained the models.
The key 
finding was that the removal of none of the feature groups could improve
the accuracy 
of all models on all platforms. For example, while the removal of
path pattern features 
does not affect accuracies of models on Last.fm, it can do so
on Quora by around $9\%$ (from $\simeq60\%$ to $\simeq51\%$).
Therefore, for the
sake of consistency 
we keep the set of features the same on both platforms. Details of
feature group removal are presented in Table~\ref{tab:ablation}.
To systematically study variations in features, we tested the effects of
adding/removing/replacing single features. 
For instance, to compute the aggregate similarity of the feed item to 
the explanation path, we plugged in two different similarity functions: 
embedding-based similarity and 
graph-based similarity. In the former, we computed the cosine 
similarity between the embeddings of categories/tags. To learn the 
embedding of each category/tag, we first sampled a set of related sentences. 
On Quora, we sampled $100$ questions at random, asked in the concerned category. 
For Last.fm tags, we treated each tag as a sentence. Then we learned the 
embedding of each sentence using the 
latent variable generative model in 
Arora et al.~\citep{arora2016simple} and represented the category/tag with the 
average embedding of the sampled 
sentences.
In the graph-based similarity function, we used the taxonomic distance 
between categories/tags in the category DAG. Replacing the graph-based
similarity with embeddings 
improved the accuracy by $8$ percent (from $\simeq52\%$ to $\simeq60\%$).
This emphasizes the 
insufficiency of graph structures in capturing the similarity of nodes.
We tried several other variations too. For example, adding counts of
\textit{node types}
as a new feature, or replacing total edge counts
with edge type counts  
(user-content, user-user and content-content) counts,
or replacing all 
user activity features with their maximum values
(instead of averages):
all of these hurt the accuracy of 
the Quora relevance model ($\simeq60\%$) by at least $\simeq0.05\%$. 

\begin{table}[t] 
	\resizebox{\columnwidth}{!}{
	\begin{tabular}{l c c c c c}
		\toprule
		\multirow{2}{*}{\textbf{Sampling}}& \multicolumn{2}{c}{\textbf{Quora}} 					
		&&  \multicolumn{2}{c}{\textbf{Last.fm}}\\
		\cmidrule{2-3}\cmidrule{5-6}
		& \textbf{Relevance} & \textbf{Surprisal} && \textbf{Relevance} & \textbf{Surprisal} \\ \toprule					
		Random 				& $\boldsymbol{60.33}$ 		& $\boldsymbol{60.38}$ 	& & $56.24$ 			& $54.21$ 				\\
		Perturb user 		& $56.36$ 					& $56.66$ 				& & $\textbf{57.77}$ 	& $\boldsymbol{58.57}$ \\
		Perturb category 	& $58.51$ 					& $58.25$ 				& & $55.52$ 			& $56.84$ 				\\
		Perturb item		& $50.26$ 					& $50.35$ 				& & $52.85$ 			& $52.14$ 				\\ \bottomrule
	\end{tabular}}
	\caption{Effect of sampling strategy on FAIRY performance. The highest value
	in a column is marked in bold.}
	\label{tab:perturb}
	\vspace*{-1cm}
\end{table}

\textbf{Perturbation analysis}. To understand the effect of instances of
each node type 
on users' judgments, we changed our sampling strategy so that paths
in each pair 
differ in only one instance. In other words, one path can be obtained by 
perturbing any one instance of the other path. For example, Alice 
$\xrightarrow{follows}$ Bob $\xrightarrow{follows}$ Health is a
user-perturbation 
of:
Alice $\xrightarrow{follows}$ \textbf{Jack} $\xrightarrow{follows}$ Health.
We generated perturbed paths for each node type and retrained the LTR models. 
Table~\ref{tab:perturb} shows that in most cases, random 
sampling fared better than such perturbed sampling.
The only exception 
is for user-perturbed paths in Last.fm. In one-on-one interviews with 
users at the end of the study, some users 
explained their preferences (more relevant) as being caused by 
presence of some specific friends on the paths. User-perturbation
created pairs of paths where such friends were present and absent, resulting
in clearer preferences and improved modeling.
The last 
row of the table, however, shows that
item-perturbations greatly degraded performance. This can be attributed to
incomplete knowledge of the users on certain path items
in the interaction graphs: replacing such items by
yet other unfamiliar items clearly has an arbitrary effect on
assessments, which seemed to worsen model accuracies.

\noindent \textbf{Transitivity of judgments.} Relevance and surprisal
are subtle factors, and it is worthwhile to investigate transitivity
in users'
assessments (for both understanding and as a sanity check).
So we extracted all triplets $(r_i, r_j, r_k)$ of explanation paths 
where we had the users' judgments on any two \textit{pairs}. We then 
computed a
$transitivity\!-\!{score} =\frac{\sum_{(r_i, r_j, r_k)}I(r_i, r_j, r_k)}
{\sum_{(r_i, r_j, r_k)}\mathbf{1}}$ 
where $I(r_i, r_j, r_k)$ is the indicator of a transitive triplet.
On a positive note, it turned out that people were transitive
(consistent)
in their judgments for $80\%$
and $74\%$ of the Quora and Last.fm triplets respectively.

\noindent \textbf{Surprisal and complexity.} To make sure that users were not 
simply using complexity (say, length) of a path as a proxy for surprisal,
we asked about a third of the users to additionally select the 
``more complex'' path. We noticed that for $\simeq7.5\%$ of pairs, the more 
surprising path was in fact the simpler (shorter) 
one, indicating that surprisal is based on more implicit factors.


\begin{table*} [t]
	\centering
	\resizebox{\textwidth}{!}{
	\begin{tabular}{c c c c c c c}
		\toprule		
		\multirow{4}{*}{\rotatebox[origin=c]{90}{Quora: Relevance}} & \multirow{2}{*}{1} & \multicolumn{5}{l}{{\color{blue}$Shrey \xrightarrow{follows} Ali \xrightarrow{follows} \textit{Social Psychology} \xrightarrow{\textit{belongs to}^{-1}} \textit{What are the things you shouldnt say to people who hate themselves}$}} \\
		 & & \multicolumn{5}{l}{${\color{blue}Shrey \xrightarrow{follows} Business \xrightarrow{\textit{belongs to}} \textit{Advice on working with people} \xrightarrow{\textit{belongs to}} \textit{Social psychology} \xrightarrow{\textit{belongs to}^{-1}} \textit{What are the things ...}}$} \\
		
		& \multirow{2}{*}{2} & \multicolumn{5}{l}{${\color{red}Ali \xrightarrow{follows} \textit{Travel} \xrightarrow{\textit{follows}^{-1}} Stephanie \xrightarrow{asks} \textit{What is the tackiest thing you have ever seen at a wedding?}}$} \\
		 & & \multicolumn{5}{l}{${\color{red}Ali \xrightarrow{follows} \textit{What is something you have tried but will never do again} \xrightarrow{\textit{belongs to}} \textit{Experiences in Life}	\xrightarrow{\textit{belongs to}^{-1}} \textit{What is the ...}}$} \\
		\midrule
		\multirow{4}{*}{\rotatebox[origin=c]{90}{Quora: Surprisal}} & \multirow{2}{*}{3} & \multicolumn{5}{l}{${\color{blue}Amr \xrightarrow{follows} Cooking \xrightarrow{\textit{follows}^{-1}} Ratnesh \xrightarrow{asks} \textit{How can you learn faster} \xrightarrow{\textit{answers}^{-1}} \textit{answer by Ara}}$} \\
		& & \multicolumn{5}{l}{${\color{blue}Amr \xrightarrow{upvotes} \textit{answer by James} \xrightarrow{answers} \textit{How can you learn faster} \xrightarrow{\textit{answers}^{-1}} \textit{answer by Ara}}$} \\
	
		& \multirow{2}{*}{4} & \multicolumn{5}{l}{${\color{red}Ali \xrightarrow{upvotes} \textit{answer by Gerry} \xrightarrow{answers} \textit{...something you secretly regret} \xrightarrow{\textit{belongs to}} \textit{life and living} \xrightarrow{\textit{belongs to}^{-1}} \textit{How do I give up on life}}$} \\
		& & \multicolumn{5}{l}{${\color{red}Ali \xrightarrow{follows} Health \xrightarrow{\textit{belongs to}^{-1}} swimwear \xrightarrow{\textit{belongs to}} \textit{life and living} \xrightarrow{\textit{belongs to}^{-1}} \textit{How do I give up on life}}$} \\
		\midrule
		
		\multirow{4}{*}{\rotatebox[origin=l]{90}{Last: Relevance}} & \multirow{2}{*}{5} & \multicolumn{5}{l}{${\color{blue}Sahar \xrightarrow{scrobbles} Earth \xrightarrow{sings^{-1}} \textit{Sleeping at last} \xrightarrow{\textit{belongs to}} \textit{indie rock} \xrightarrow{\textit{belongs to}^{-1}} artist:Kodaline}$} \\
		& & \multicolumn{5}{l}{${\color{blue}Sahar \xrightarrow{follows} Sana \xrightarrow{scrobbles} \textit{Adventure of a lifetime} \xrightarrow{\textit{sings}^{-1}} Coldplay \xrightarrow{\textit{belongs to}} alternative \xrightarrow{\textit{belongs to}^{-1}} artist:Kodaline}$} \\
				
		&\multirow{2}{*}{6} & \multicolumn{5}{l}{${\color{red}Bahar \xrightarrow{follows}	Mojtaba	\xrightarrow{scrobbles}	Baribakh \xrightarrow{sings^{-1}} Mansour \xrightarrow{\textit{belongs to}}	Persian \xrightarrow{\textit{belongs to}^{-1}} artist:Ebi		}$} \\
		& & \multicolumn{5}{l}{${\color{red}Bahar \xrightarrow{\textit{loves}} \textit{Twist in my sobriety} \xrightarrow{\textit{belongs to}}	pop	\xrightarrow{\textit{belongs to}^{-1}} track:Pickack	\xrightarrow{sings}	artist:Ebi}$} \\
		
		\midrule
		\multirow{4}{*}{\rotatebox[origin=c]{90}{Last: Surprisal}} & \multirow{2}{*}{7} & \multicolumn{5}{l}{${\color{blue}Elitsa \xrightarrow{follows} Engekbert \xrightarrow{scrobbles} \textit{The prince} \xrightarrow{\textit{belongs to}} metal \xrightarrow{\textit{belongs to}^{-1}} track:Rigger \xrightarrow{\textit{sings}^{-1}} artist: In flames}$} \\
		& & \multicolumn{5}{l}{${\color{blue}Elitsa \xrightarrow{loves} \textit{Ohne dich} \xrightarrow{contains^{-1}} \textit{Reise, Reise} \xrightarrow{\textit{belongs to}} metal \xrightarrow{\textit{belongs to}^{-1}} \textit{track : Acoustic piece} \xrightarrow{\textit{sings}^{-1}}
			artist:In flames		}$} \\				
		
		&\multirow{2}{*}{8}& \multicolumn{5}{l}{${\color{red}Ali \xrightarrow{loves} \textit{bang bang}	\xrightarrow{\textit{belongs to}} \textit{female vocalists} \xrightarrow{\textit{belongs to}^{-1}} \textit{album : Dua Lipa (Deluxe)} \xrightarrow{contains} Dreams \xrightarrow{\textit{sings}^{-1}}	\textit{Dua lipa}}$} \\
		& & \multicolumn{5}{l}{${\color{red}Ali \xrightarrow{follows} Anna \xrightarrow{scrobbles} \textit{bad girl friend} \xrightarrow{\textit{belongs to}} pop \xrightarrow{\textit{belongs to}^{-1}} \textit{track : New rules} \xrightarrow{sings^{-1}} \textit{Dua lipa}}$} \\
		
		\bottomrule
	\end{tabular}}
	\caption{Anecdotal cases of path pairs with id's. FAIRY makes correct
	and wrong predictions for
	blue and red pairs, respectively.}
	\label{tab:anecdotes}
	\vspace*{-0.7cm}
\end{table*}

\textbf{Anecdotal examples}. Table~\ref{tab:anecdotes} presents some examples 
of correct (in \textcolor{blue}{blue}) and wrong (in \textcolor{red}{red})
predictions
by FAIRY. 
In each pair, the first path denotes the preferred one by the user.
The diversity of node and edge types in correct pairs shows the ability of
FAIRY to 
learn underlying factors determining relevance or surprisal.
The wrongly predicted pairs provide
insights on the shortcomings of FAIRY. For example, in the surprisal model,
we do not consider \textit{sensitivity} of topics or items.
Pair \#4 
is one such example where explanation 
paths reveal such sensitive items the users interacted with. 
Another limitation of FAIRY is that it does not incorporate background 
information about the users (such as their nationality) which clearly affects 
their preferences. 
For example, on Last.fm,
some users mentioned that presence of ethnic tags
(such as Persian, Latin or Brazilian) influenced their relevance decisions
(pair \#6).



	
		 

%% file: sections/related-work.tex
\section{Related Work} 
\label{sec:related}

\noindent \textbf{Social feeds and transparency.}
Personalizing social feeds has been the focus of many studies, as the 
amount of information generated by users' networks is overwhelming.
To increase user engagement, models have been developed 
for finding relevant feed items for users by exploiting their past behavior
~\cite{freyne2010social,hong2012learning,soh2013recommendation,
agarwal2015personalizing}. Users, however, are often unaware of the presence
of such curation algorithms~\cite{hamilton2014path,eslami2015always}
as service providers generally 
do not provide insightful explanations. For example,
Cotter et al.~\cite{cotter2017explaining} demonstrate 
inadequacy of explanations for feed ranking in Facebook. 

\noindent \textbf{Heterogeneous information networks and meta-paths.}
Due to limitations of traditional
graphs for capturing
complex semantics 
with different types of entities and relations,
heterogeneous information networks (HIN) 
were introduced to model multiple node and edge types~\citep{sun2013mining}. 
To analyze such HINs better,
a meta-path was defined as the pattern of a path (sequence of node and
edge types). 
Meta-paths have since been used in different applications 
such as 
similarity search~\cite{sun2011pathsim,seyler2018information},
relationship discovery~\cite{fang2011rex,kong2013multi,behrens2018metaexp}, 
link prediction~\cite{sun2012will,zhang2014meta,zhang2018camel,dong2017metapath2vec},
and generating recommendations~\cite{lee2013pathrank,hu2018leveraging,liu2014meta} 
in HINs. 

\noindent \textbf{Relationship discovery in knowledge graphs.}
Finding interesting relationships among graph concepts
is too broad an area to do justice in a short survey: however,
mining such connections for knowledge graph entities
is a more pertinent sub-problem that has been well-studied.
This task is either (semi-)supervised,
where users are asked for feedback~\cite{behrens2018metaexp},
or unsupervised, where heuristic measures of
``interestingness''
are applied to detect and rank relationships. However, user utility is probably
multi-faceted: while we have explored relevance and surprisal,
there are probably more, like coherence or complexity. These measures 
are normally approximated using topological properties of
graphs, such as specificity and rarity of node/edge/path types 
~\cite{ramakrishnan2005discovering,fang2011rex} or connectivity/reachability of 
nodes~\cite{lao2010relational,seufert2016espresso,liang2016links}.  
These scoring strategies, however,
implicitly assume a static topology, and might not be very useful for 
dynamic interaction graphs where nodes and edges are added
and deleted
with each timestep, and it is imperative that relationships
should take temporal constraints into account.

%% file: sections/conclusion.tex
\vspace*{-0.1cm}

\section{Conclusion}
\label{sec:conclusions}

We presented FAIRY, a smart user-side framework that presents
ranked explanations to users for items in their social feeds. Explanations are
represented as relationship paths connecting the user's own actions
to the received feed items.
FAIRY was trained and evaluated on data from two real user studies
on the popular platforms of Quora and Last.fm. It outperformed
three baselines on relationship mining on the task of modeling and predicting
what users considered relevant and surprising explanations.
The success of FAIRY hinges on two key aspects: a powerful
heterogeneous information network representation of the
user's local neighborhood that can capture the complexity of current social
media platforms, and, (ii) a fast learning-to-rank model that operates with
intuitive and interpretable features that are easily accessible to the user.

FAIRY is the first step towards a general goal of
improving transparency through the user's lens.
We plan to implement FAIRY
as a browser plug-in for users.
Future directions for research would include,
among others: (i) better modeling of temporal information in the interaction
graph, (ii) further exploiting content-features to build better models of
entity similarity, and (iii) understanding effects of the user's activities
across multiple connected platforms.

\noindent \textbf{Acknowledgements.} 
This work was partly supported by the ERC
Synergy Grant 610150 (imPACT) and the DFG Collaborative Research 
Center 1223. We would like to thank Krishna P. Gummadi from MPI-SWS 
for useful discussions at various stages of this work.